\documentclass[preprint,12pt]{article}
\usepackage{graphicx}
\usepackage[margin=1in]{geometry}
\usepackage{amssymb}
\usepackage[labelfont=bf]{caption}
\usepackage{subcaption}
\usepackage{amsmath} 

\begin{document}

\title{Doping evolution of the Mott-Hubbard landscape in infinite-layer nickelates}

\author{Berit H. Goodge$^{1,2}$, Danfeng Li$^{3,4}$, Motoki Osada$^{3,5}$, \\ Bai Yang Wang$^{3,6}$, Kyuho Lee$^{3,6}$, George A. Sawatzky$^{7,8}$, \\ Harold Y. Hwang$^{3,4}$, \& Lena F. Kourkoutis$^{1,2}$*}

\date{
\normalsize{$^{1}$School of Applied and Engineering Physics, Cornell University, Ithaca, NY 14853, USA}
\\
\normalsize{$^{2}$Kavli Institute at Cornell for Nanoscale Science, Cornell University, Ithaca, NY 14853, USA}
\\
\normalsize{$^{3}$Stanford Institute for Materials and Energy Sciences, SLAC National Accelerator Laboratory,}
\\
\normalsize{Menlo Park, CA 94025, USA}
\\
\normalsize{$^{4}$Department of Applied Physics, Stanford University, Stanford, CA 94305, USA}
\\
\normalsize{$^{5}$Department of Materials Science and Engineering, Stanford University, Stanford, CA 94305, USA}
\\
\normalsize{$^{6}$Department of Physics, Stanford University, Stanford, CA 94305, USA}
\\
\normalsize{$^{7}$Department of Physics and Astronomy, University of British Columbia, }
\\
\normalsize{Vancouver, B.C. V6T 1Z1, Canada  }
\\
\normalsize{$^{8}$Stewart Blusson Quantum Matter Institute, University of British Columbia,}
\\
\normalsize{Vancouver, B.C. V6T 1Z1, Canada}
\\
\normalsize{$^\ast$To whom correspondence should be addressed; email: lena.f.kourkoutis@cornell.edu.}
 }

\maketitle

\textbf{
The recent observation of superconductivity in Nd$_{0.8}$Sr$_{0.2}$NiO$_2$ has raised fundamental questions about the hierarchy of the underlying electronic structure. Calculations suggest that this system falls in the Mott-Hubbard regime, rather than the charge-transfer configuration of other nickel oxides and the superconducting cuprates. Here, we use state-of-the-art, locally-resolved electron energy loss spectroscopy to directly probe the Mott-Hubbard character of Nd$_{1-x}$Sr$_x$NiO$_2$. Upon doping, we observe emergent hybridization reminiscent of the Zhang-Rice singlet via the oxygen-projected states, modification of the Nd 5$d$ states, and the systematic evolution of Ni 3$d$ hybridization and filling. These results clearly evidence the multiband nature of this system and the distinct electronic landscape for infinite-layer nickelates despite their formal similarity to the cuprates.}
\newpage

The discovery of copper oxide high-temperature superconductors \cite{bednorz_possible_1986} and subsequent realizations of non-copper based compounds \cite{maeno_superconductivity_1994, kamihara_iron-based_2008} has spurred significant efforts to not only understand the mechanisms underlying superconductivity \cite{armitage_progress_2010, hosono_iron-based_2015, keimer_quantum_2015, mackenzie_even_2017}, but also to identify and realize new host materials. 
The recent discovery of superconducting nickel-based thin films \cite{li_superconductivity_2019} has opened an opportunity to explore a system which is closely related to the infinite-layer cuprates in both crystal structure and transition metal electron count. Trading out the Cu$^{2+}$ ions for Ni$^{1+}$ preserves an isoelectronic formal 3$d^9$ state, raising the possibility of a nickel analogue to the cuprates \cite{anisimov_electronic_1999}. At the same time, key differences between the systems have been proposed both early on \cite{lee_infinite-layer_2004} and more recently following the successful demonstration of nickelate superconductivity \cite{nomura_formation_2019,hu_two-band_2019,choi_role_2020, hirayama_materials_2020, jiang_doped_2019,  botana_similarities_2019}. Experimentally, however, direct measurements of the electronic landscape of hole-doped infinite-layer nickelates, particularly with variable band filling, are lacking.

In the family of superconducting cuprates, extensive spectroscopic characterisation of bulk samples enabled a thorough understanding of their electronic structure, especially regarding the role of doped holes \cite{keimer_quantum_2015}. Compared to such bulk phase systems, the successful realization of nickelate superconductivity in epitaxial thin films complicates their characterization by traditional bulk techniques, instead requiring local probes as well as measurements with high sensitivity due to the metastable nature of these compounds. Here, we directly probe the electronic structure across the doped Nd$_{1-x}$Sr$_x$NiO$_y$ ($y$ = 2, 3) family using locally-resolved electron energy loss spectroscopy (EELS) in the scanning transmission electron microscope (STEM) highly optimized for detection of subtle spectroscopic signatures. We identify a distinct pre-peak feature in the NdNiO$_3$ O-K edge which disappears completely in NdNiO$_2$, consistent with the predicted Mott-Hubbard character of the infinite-layer parent compound \cite{lee_infinite-layer_2004, jiang_doped_2019}. Moreover, our systematic study across a series of superconducting hole-doped Nd$_{1-x}$Sr$_x$NiO$_2$ ($x \leq$ 0.225) films uncovers an emergent feature in the O 2$p$ band with hole doping, reminiscent of the Zhang-Rice singlet peak in the isostructural superconducting cuprates. Unlike the cuprates, however, the spectral weight of the O 2$p$ feature is small even at high doping levels, suggesting necessary involvement of other bands. Indeed, parallel spectroscopy of both Ni and Nd confirms contributions of the Ni-3$d$ and modification of the Nd-5$d$ states, demonstrating, in contrast to the superconducting cuprates, the multiband nature of this system.

\begin{figure*}[t!]
    \centering
    \includegraphics[width=\linewidth]{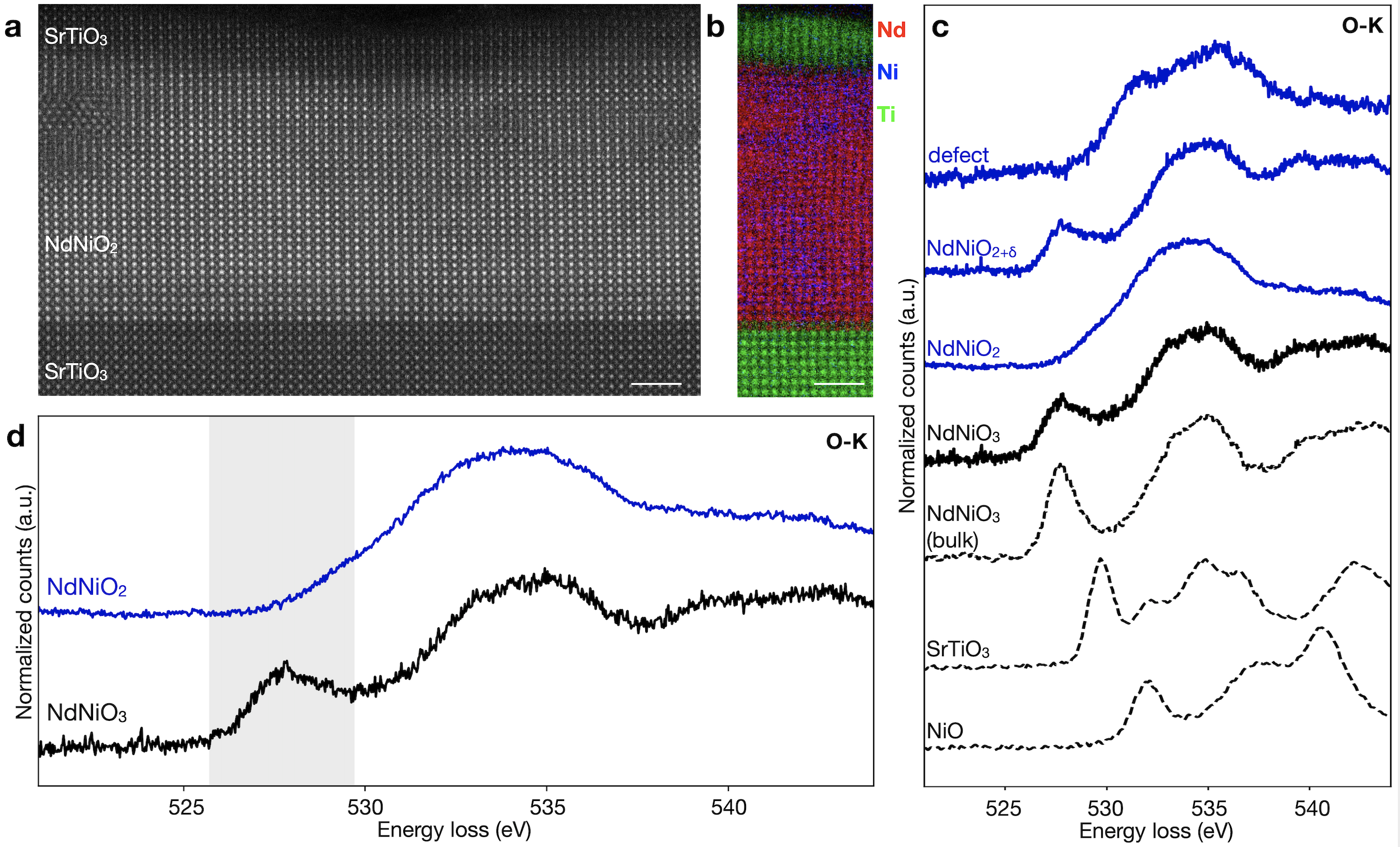}
    \caption{\textbf{Lattice and electronic structure of NdNiO$_y$ thin films.} \textbf{a}, Atomic-resolution HAADF-STEM imaging of an undoped NdNiO$_2$ film shows mostly well-ordered epitaxial structure with some visible crystalline defects. \textbf{b}, STEM-EELS Nd, Ni, and Ti elemental mapping confirms an abrupt interface with SrTiO$_3$. \textbf{c}, Observed variations in O-K near-edge fine structure across different regions in the same reduced nickelate film (blue) necessitate local measurements to avoid contributions from defects and partially unreduced regions. For comparison, reference spectra from bulk oxides are also shown. \textbf{d}, EELS O-K edge of NdNiO$_2$ and a NdNiO$_3$ film before reduction to the infinite-layer phase. The disappearance of the first peak (highlighted in grey), ascribed to metal-oxygen hybridization, indicates filling of the Ni states upon reduction from $d^7$ to $d^9$. Scale bars 2 nm.}\label{fig:variation}
\end{figure*}

Infinite-layer nickelate thin films are stabilized by soft-chemistry topotactic reduction \cite{hayward_sodium_1999} of the perovskite phase grown epitaxially on SrTiO$_3$ (001) substrates and capped with a thin layer of SrTiO$_3$ by pulsed laser deposition (PLD) as previously described \cite{li_superconductivity_2019, lee_aspects_2020}. Capping layers between 2-25 nm have been found to stabilize and support the infinite-layer structure during topotactic reduction, producing more crystallographically uniform films \cite{lee_aspects_2020}. STEM-EELS is therefore an ideal way to spectroscopically probe the nickelate thin film without contributions from the capping layer (or substrate). Starting with the undoped NdNiO$_2$ film, atomic-resolution high-angle annular dark-field (HAADF) scanning transmission electron microscopy (STEM) imaging confirms the overall high crystalline quality of the film (Fig. \ref{fig:variation}a) after optimization of the epitaxial growth and subsequent reduction process as discussed elsewhere \cite{lee_aspects_2020}. Elemental EELS mapping performed in the STEM confirms the abruptness of the nickelate-SrTiO$_3$ interface and shows no obvious impurity phases (Fig. \ref{fig:variation}b). Extended defects are, however, present and must be taken into account when interpreting more commonly used area-averaged spectroscopic measurements (Fig. S1, S2). Spatially resolved STEM-EELS reveals distinct O-K near-edge fine-structures in different regions of the same film including some crystalline defects and small pockets that have not completely reduced to NdNiO$_2$ (Fig. \ref{fig:variation}c and Fig. S2). The intrinsically metastable nature of the infinite-layer compound \cite{hayward_sodium_1999} has required an empirical fine tuning of the conditions for maximal reduction without decomposition \cite{lee_aspects_2020}, so the presence of some unreduced pockets is likely unavoidable at this time. When averaging over the entire film, however, contributions from such variations can mask the electronic character of the pure nickelate phase. In this work, we extract the electronic signatures of the Nd$_{1-x}$Sr$_x$NiO$_y$ phases without contributions from extended defects by confining our spectroscopic measurements to crystallographically clean and fully reduced regions of each film using an~\AA-size STEM probe.

We first explore the electronic character of the parent infinite-layer compound, NdNiO$_2$.  In 3$d$ transition metal oxides, the onset structure of the O-K edge is ascribed to hybridization between the O 2$p$ and metal 3$d$ states \cite{de_groot_oxygen_1989}. For NdNiO$_3$, this hybridization results in a strong pre-peak feature at $\sim$527 eV \cite{de_groot_oxygen_1989, abbate_electronic_2002, palina_investigation_2017}. Compared to the bulk compound, the pre-peak in the NdNiO$_3$ thin film is broadened, consistent with changes to local Ni-O coordination \cite{abbate_electronic_2002, gauquelin_atomically_2014} due to epitaxial strain imposed by the SrTiO$_3$ substrate (Fig. \ref{fig:variation}c). Upon oxygen reduction from the perovskite NdNiO$_3$ to the infinite-layer NdNiO$_2$ phase, the pre-peak disappears entirely (Fig. \ref{fig:variation}d). Here, the suppression of this peak reflects filling of the lower $d$ bands with the change in formal Ni configuration from 3$d^7$ to 3$d^9$ \cite{abbate_electronic_2002}.
The undoped infinite-layer cuprates exhibit a similar suppression in spectral weight of the $d^9$ states, but with an additional O-K peak due to transitions into the $d^{10}$ upper Hubbard band observed $\sim$2.4 eV higher in energy  \cite{chen_electronic_1991,meyers_zhang-rice_2013}. In the nickelates, however, a similar second peak is absent. 
Recent theoretical work has suggested that the reduction in nuclear charge from Cu to Ni greatly reduces the covalency of the $d^9$ state and pushes the $d^{10}$ states to high energies \cite{jiang_doped_2019}, such that a $d^{10}$ peak like that observed in the cuprates would be much weaker and located several ($\sim$5-6) eV higher in energy, overlapping with the rest of the O-K edge.
The lack of this peak in the nickelates therefore indicates an electronic structure with a charge-transfer gap $\Delta$ larger than the on-site Couloumb interaction $U$. Using the Zaanen-Sawatzky-Allen (ZSA) classification scheme \cite{zaanen_band_1985}, this places the parent infinite-layer compound in the Mott-Hubbard regime ($\Delta > U$), in contrast to the charge-transfer ($\Delta < U$) cuprates \cite{chen_electronic_1991} (Fig. S3). Previous results by x-ray absorption spectroscopy (XAS) have similarly suggested the Mott-Hubbard nature of La-based nickelates \cite{hepting_electronic_2020}. Note that hybridization in the nickelates spreads out the distribution of the O 2$p$ bands \cite{bisogni2016ground}, likely moving the system near the boundary between the two regimes (Mott-Hubbard and charge-transfer) \cite{zaanen_band_1985}.

\begin{figure*}[b!]
    \centering
    \includegraphics[width=\linewidth]{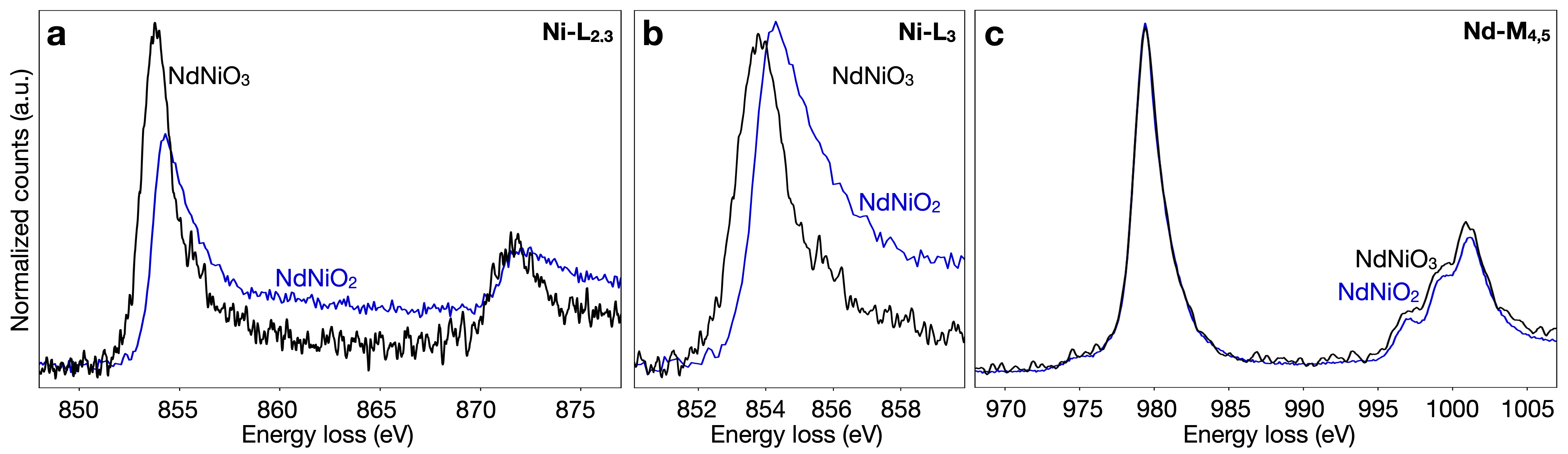}
    \caption{\textbf{Electronic evolution from perovskite NdNiO$_3$ (black) to infinite-layer NdNiO$_2$ (blue) phase for epitaxial thin films on SrTiO$_3$.} \textbf{a}, The Ni-L$_{2,3}$ edge (3$d$ states) shows a clear change upon O reduction from NdNiO$_3$ to NdNiO$_2$, including \textbf{b}, significant broadening and shift of the L$_3$ edge. \textbf{c}, The Nd-M$_{4,5}$ edge (4$f$ states) shows little or no change between the two compounds. All spectra are normalized by integrated signal over the full energy ranges shown.}\label{fig:113-112}
\end{figure*}

XAS has also been used to probe the O-K near-edge fine-structure of reduced NdNiO$_2$ and unreduced NdNiO$_3$ films on SrTiO$_3$ \cite{hepting_electronic_2020,karp2020manybody}, but these measurements averaged over large parts of the films. The exclusion of defects and the stabilizing SrTiO$_3$ capping layer in our spatially resolved approach explains the apparent discrepancy between these results. In particular, contributions from SrTiO$_3$ or secondary phases in non-optimized films \cite{lee_aspects_2020} result in an additional peak at $\sim$530 eV in the unreduced film as well as a subtle but clear shoulder at the same energy in the reduced film (Fig. S1). While differences in the perovskite phase are quite striking, below we show that even small contributions from this shoulder in the infinite-layer films can inhibit the ability to probe the effects of hole doping in this system.

In addition to the O-K edge, EELS measurements provide simultaneous access to the Ni and Nd states in corresponding localized regions. We measure the Ni-L$_{2,3}$ edge (Fig. \ref{fig:113-112}a) to probe the Ni 3$d$ states \cite{muller_connections_1998} across the same evolution from perovskite to infinite-layer phase. A clear distinction can be made between the parent NdNiO$_3$ and infinite-layer NdNiO$_2$ thin films, particularly by a broadening and shift of the L$_3$ edge at $\sim$854 eV (Fig. \ref{fig:113-112}b). Transition metal L$_{2,3}$ edges have been used to fingerprint valence states in more ionic systems. In nickel compounds, however, hybridization effects play an important role and are reflected in the fine structure of the Ni-L$_{2,3}$ edge \cite{muller_connections_1998}. Similar to the O-K edge, the spectral differences between bulk and thin film NdNiO$_3$ in the L$_{2,3}$ are likely due to strain effects \cite{kim_strain_2020}. Comparing the two thin film phases, the asymmetric broadening and long high-energy tails of the Ni-L$_3$ edge in the reduced NdNiO$_2$ suggests increased $p$-$d$ hybridization in the infinite-layer phase \cite{muller_connections_1998,koyama_electronic_2005,palina_investigation_2017}. Unlike the Ni-L$_{2,3}$ and O-K, the Nd-M$_{4,5}$ edge (4$f$ states) remains unchanged upon reduction from perovskite to infinite-layer (Fig. \ref{fig:113-112}c), indicating a formal Nd valence of 3+ in both phases (Fig. S4).

\begin{figure*}[b!]
    \centering
        \includegraphics[width=\linewidth]{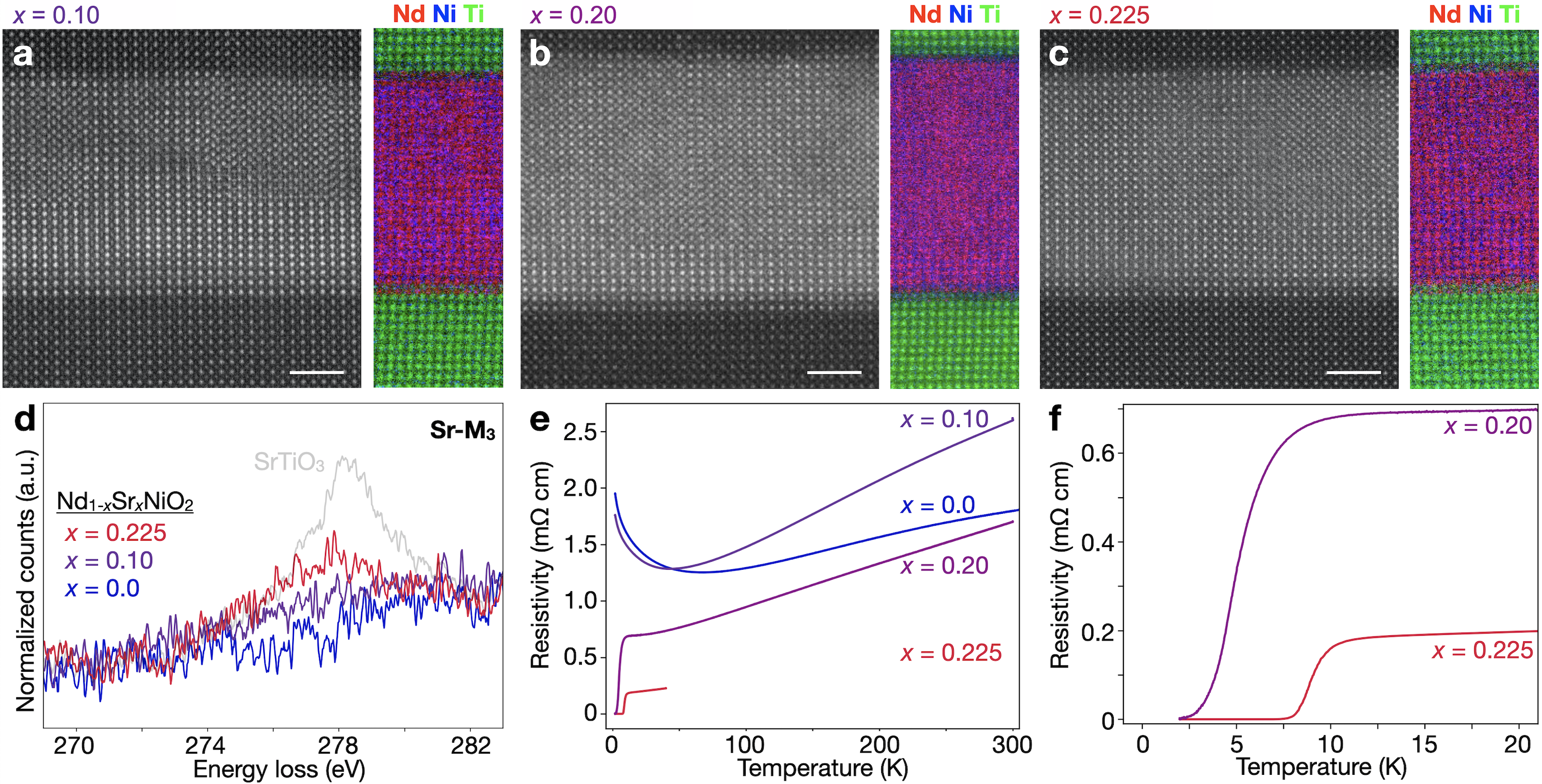}
    \caption{\textbf{Structure and transport of Nd$_{1-x}$Sr$_x$NiO$_2$ films with hole doping $x \leq$ 0.225.} \textbf{a-c}, Atomic resolution HAADF-STEM and elemental EELS maps of $x$ = 0.10, 0.20 and 0.225 samples, respectively. \textbf{d}, Sr-M$_3$ EEL spectra from each film show a qualitative trend consistent with the expected nominal Sr-doping in each sample. A similar spectrum from SrTiO$_3$ is shown in grey for comparison. \textbf{e}, \textbf{f}, Resistivity vs. temperature curves for the four Nd$_{1-x}$Sr$_x$NiO$_2$ films studied here. Lightly Sr-doped ($x$ = 0.0, 0.1) films show metallic behavior with a resistive upturn near 50-70 K, while more strongly doped ($x$ = 0.2, 0.225) films show superconducting transitions (T$_{c,90\%R}$) at 8 and 11 K, respectively. Data for $x$ = 0.0 from \cite{li_superconductivity_2019}.}\label{fig:doping_struc}
\end{figure*}

Having established the nature of the parent compounds, we systematically study the effect of hole doping in a series of superconducting infinite-layer Nd$_{1-x}$Sr$_x$NiO$_2$ films. HAADF-STEM imaging and elemental EELS mapping (Fig. \ref{fig:doping_struc}a-c) confirm overall film quality across all samples with uniform distribution of Nd and Ni throughout the films. Summed spectra (Fig. \ref{fig:doping_struc}d) from clean regions within the films show Sr-M$_3$ signals consistent with the relative nominal Sr doping expected in each case. Transport measurements for the infinite-layer films studied here (Fig. \ref{fig:doping_struc}e, f) show metallic temperature dependence of the low ($x$ = 0.0, 0.1) Sr (hole) doped films down to about 50-70 K, below which a resistive upturn is observed. Upon increased hole doping ($x$ = 0.2, 0.225), the infinite-layer films become superconducting with transition temperatures $T_{\textrm{c,90\%R}}$ (the temperatures at which the resistivity is 90\% that at 20 K) of 8 and 11 K, respectively (Fig. \ref{fig:doping_struc}f). The relative values of $T_\textrm{c}$ for these films are consistent with a recent investigation of the doping dependence of Nd$_{1-x}$Sr$_x$NiO$_2$ establishing a superconducting dome spanning 0.125 $< x <$ 0.25 \cite{denver_new}.

\begin{figure*}
    \centering
        \includegraphics[width=0.8\linewidth]{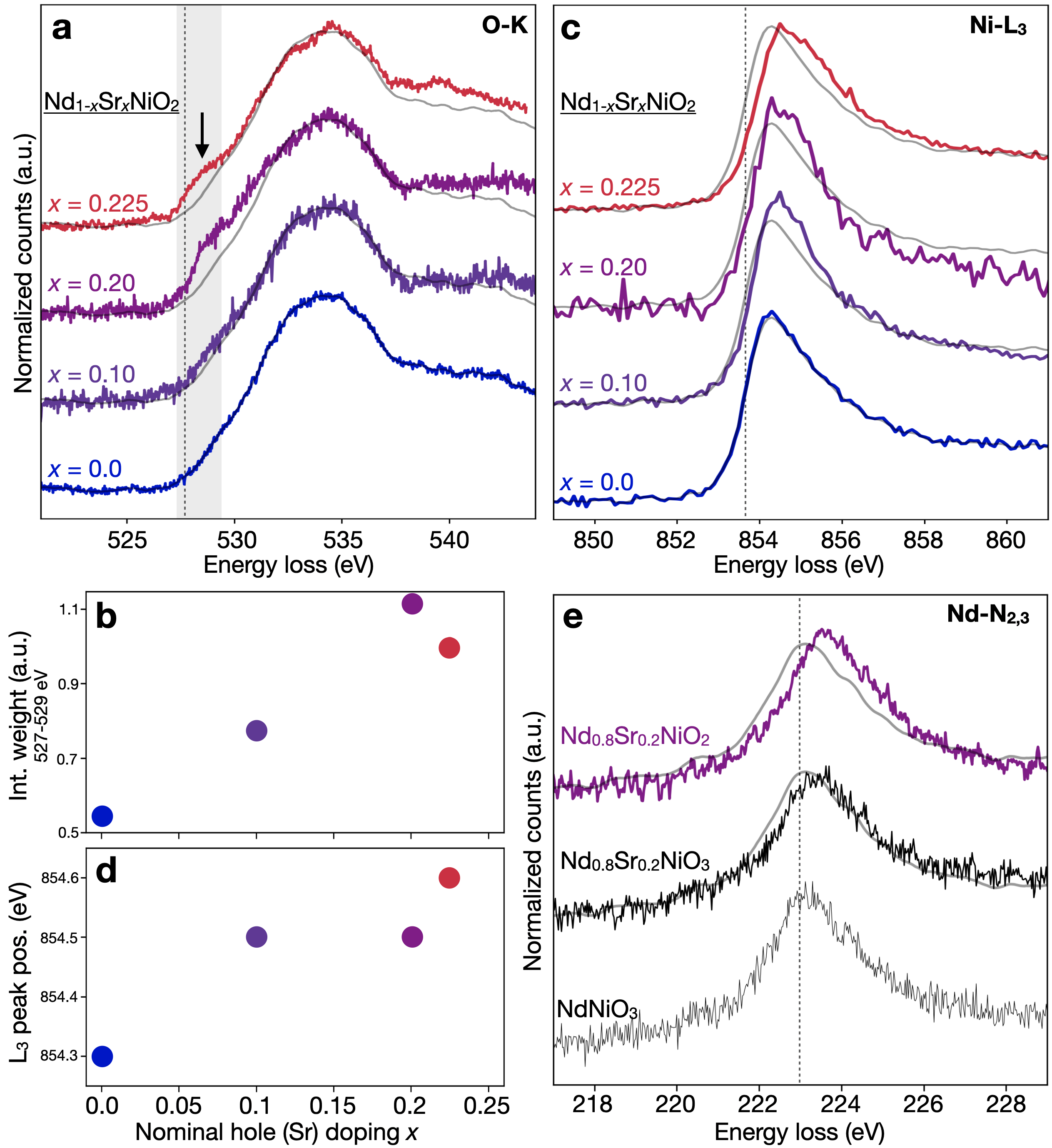}
    \caption{\textbf{Electronic structure evolution of Nd$_{1-x}$Sr$_x$NiO$_2$ with hole doping $x \leq$ 0.225.} \textbf{a}, A spectral feature at $\sim$528 eV emerges with increased Sr (hole) doping in the O-K edge, attributed to $d^9\underline{L}$ states. The dotted line marks the $d^9$ peak position in NdNiO$_3$. \textbf{b}, The integrated signal over the $d^9\underline{L}$ peak spectral range (527-529 eV, highlighted grey in \textbf{a} increases with hole doping. \textbf{c}, \textbf{d}, The Ni-L$_3$ edge systematically shifts to high energies and broadens with hole-doping. The dotted line marks the L$_3$ peak position in NdNiO$_3$. \textbf{e}, The Nd-N$_{2,3}$ edge (5$d$ states) shows a small shift to higher energies upon Sr-doping of the perovskite phase. A further shift is observed upon O-reduction to infinite-layer Nd$_{0.8}$Sr$_{0.2}$NiO$_2$. The dotted line marks the N$_{2,3}$ peak position in NdNiO$_3$.
    For comparison, reference spectra are plotted in grey for NdNiO$_2$ in \textbf{a}, \textbf{c} and for NdNiO$_3$ in \textbf{e}.}\label{fig:doping_elec}
\end{figure*}

In the O-K edge, we observe an emergent feature at $\sim$528 eV as a function of Sr (hole) doping across the Nd$_{1-x}$Sr$_x$NiO$_2$ films (Fig. \ref{fig:doping_elec}a).  The strength of this peak increases with hole doping (Fig. \ref{fig:doping_elec}b), $x$, although the spectral weight of the $x$ = 0.2 sample is higher than the more strongly doped $x$ = 0.225 sample, likely due to the inclusion of some RP-fault regions (Figs. S1, S5). A similar spectral feature is also observed upon hole doping in the cuprates \cite{kuiper_x-ray_1988, chen_electronic_1991}. As in the cuprates, we attribute the emergent spectral feature to doped holes in the oxygen orbitals forming 3$d^9\underline{L}$ states, where $\underline{L}$ is the oxygen ``ligand'' hole \cite{kuiper_x-ray_1988, armitage_progress_2010}. The coordination of two such holes into a Zhang-Rice singlet (ZRS) state is thought to reduce the cuprate system into an effective single-band model \cite{zhang_effective_1988, armitage_progress_2010}. In nickelates the picture is less clear: despite the similarity in presence and doping-dependence of this O-K edge peak in both systems, a few key spectroscopic differences are apparent. In the cuprates, hole doping results in a tradeoff of spectral weight from the $d^{10}$ upper Hubbard band into the emergent $d^{9}\underline{L}$ states \cite{chen_electronic_1991}. In the nickelates, we do not observe a similar trend, most likely due to shift of the $d^{10}$ band to much higher energy as described by the Mott-Hubbard picture discussed above. Perhaps more importantly, the overall strength of the hole doping peak (3$d^9\underline{L}$) is significantly reduced in the nickelates compared to the cuprates, even for comparatively high levels of hole doping. The weak effect of doping on the O-K edge begs the question: where else do the holes go? 

The ability of EELS measurements to access several edges simultaneously allows us to consider the corresponding effects on the Ni-L$_{2,3}$ edge (3$d$ bands) (Fig. \ref{fig:doping_elec}c). The Ni-L$_3$ peak position (Fig. \ref{fig:doping_elec}d and Fig. S6) shifts systematically to higher energies upon hole doping, with a total increase of $\sim$0.3 eV between the undoped $x$ = 0.0 and the $x$ = 0.225 films. We also observe L$_3$ edge broadening as an extended tail in the doped samples (Fig. S5). Together, the changes in the Ni-L$_{2,3}$ edge may suggest increased hybridization and a mixed-valent Ni state. 

We also probe the Nd-N$_{2,3}$ edge (5$d$ states) across the sample series (Fig. \ref{fig:doping_elec}e). Given the extreme technical challenges due to the need for simultaneous high-spatial resolution, high-energy resolution, and high signal-to-noise ratio, current data is limited only to the compositions shown in Fig. \ref{fig:doping_elec}e, although characterizing the full doping dependence of the N$_{2,3}$ edge is the subject of ongoing work. The $x$ = 0.2 hole-doped perovskite shows a $\sim$200 meV shift to higher energies compared to the parent compound. In the $x$ = 0.2 infinite-layer film, we observe a twofold increase in this energy shift as well as an increase in spectral intensity. Given the ongoing theoretical debate about the importance of parent cation 5$d$ bands \cite{lee_infinite-layer_2004, botana_layered_2018,botana_similarities_2019}, direct experimental measurements of these states are important. Full interpretation of the observed shifts in the Nd-N$_{2,3}$ edge will, however, require detailed calculations.

Together, the evolution of all three edges (O-K, Ni-L$_{2,3}$, and Nd-N$_{2,3}$) suggest markedly different behavior of doped holes in infinite-layer nickelates than in the cuprates. Rather than doping mostly onto O, our data are consistent with a picture in which the large charge-transfer energy $\Delta > U$ of the Mott-Hubbard regime instead pushes holes also onto Ni, resulting in $d^8$ like states above the O 2$p$ band. Hybridization between the ligand holes and $d^8$ states results in the observed $d^9\underline{L}$ peak at $\sim$528 eV in the O-K edge. Hybridization with this $d^{9}\underline{L}$ pushes the $d^8$ state up towards the Fermi energy, resulting in a ``mixed-valent metal” state, as also suggested by the observed broadening of the Ni-L$_{2,3}$ edge. Finally, the effects measured in the Nd-N$_{2,3}$ edge illustrate some interaction with the doped holes in the Nd 5$d$ bands as well, suggesting a multi-band system fundamentally different from the effective single-band superconducting cuprates \cite{hu_two-band_2019, denver_new}. 

Beyond hole doping, the large parameter space (rare earth cation, epitaxial strain, reduction process, etc.) yet to be explored for this new family of superconductors will provide a rich backdrop for future experiments. Especially in thin film form, spatially resolved spectroscopy will be a powerful technique to not only minimize spectral signatures due to defects, but also exclude contributions from stabilizing capping layers \cite{lee_aspects_2020}.

The data presented here are direct evidence of key differences between the electronic structures of two isostructural superconducting oxide families. Analysis of the parent perovskite and infinite-layer compounds establishes the Mott-Hubbard character of NdNiO$_2$, distinct from the charge-transfer cuprates. Within the doped infinite-layer series, we observe effects of hole-doping in not just the O 2$p$ but also the Ni 3$d$ and Nd 5$d$ bands, drawing a further distinction between the nickelates and their isostructural cuprate cousins. 
Lacking both the charge transfer character and magnetic order \cite{hayward_synthesis_2003} of the cuprates, understanding the nature of superconductivity in this new family of nickelates will require new insights from both theory and experiment.

\section*{Materials and Methods}
\subsection*{Thin film synthesis and characterization} 

Thin films of precursor perovskite nickelate Nd$_{1-x}$Sr$_x$NiO$_3$ 8-10 nm in thickness were fabricated by pulsed-laser deposition on single-crystalline (001) SrTiO$_3$ substrates. To synthesize crystalline uniform films and improve sample-to-sample reproducibility, we adopt the recently established `high-fluence’ conditions using precise laser imaging conditions \cite{lee_aspects_2020}. For all samples, a SrTiO$_3$ epitaxial capping layer was subsequently deposited using conditions described previously \cite{li_superconductivity_2019}. After the growth, an annealing-based topochemical reduction was then employed to achieve the infinite-layer phase using CaH$_2$ as reducing reagent. The annealing conditions were optimized as described previously \cite{lee_aspects_2020}. In particular, for samples with a SrTiO$_3$ cap layer of $\sim$25 nm, the reduction temperature $T_\textrm{r}$ of 280$^{\circ}$C and reduction time $t_\textrm{r}$ of 4 – 6 hours were used; for samples with a $\sim$2 nm SrTiO$_3$ cap layer, $T_\textrm{r}$ and $t_\textrm{r}$ were 260$^{\circ}$C and 1 – 3 hours, to achieve a complete transformation to the infinite-layer phase. XRD symmetric and asymmetric scans of the nickelate films were measured using a monochromated Cu K$_{\alpha1}$ ($\lambda$ = 1.5406 \AA) source. The resistivity measurements were made using Al wire-bonded contacts.

\subsection*{Atomic-resolution STEM, EELS}
Electron transparent TEM samples of each film were prepared on a Thermo Fischer Scientific Helios G4 X focussed ion beam (FIB) using the standard liftout method. Samples were thinned to $<$30 nm with 2 kV Ga ions, followed by a final polish at 1 kV to reduce effects of surface damage. All specimens were stored in vacuum to prevent possible degradation in air. 

High-angle annular dark-field (HAADF) scanning transmission electron microscopy (STEM) was performed on an aberration-correction FEI Titan Themis at an accelerating voltage of 300 kV with a convergence angle of 30 mrad and inner and outer collection angles of 68 and 340 mrad, respectively. Elemental maps were recorded with a 965 GIF Quantum ER spectrometer and a Gatan K2 Summit direct electron detector operated in counting mode. All EELS experiments presented here were conducted with probe currents less than 20 pA to minimize radiation damage with the scan parameters used: $\sim$1 \AA\, probe, $\sim$0.1 \AA$^2$ scan pixel size, and 0.5 $\mu$s/pixel dwell time (Fig. S7).  One consequence of this low beam current is the need for long total acquisition times to achieve spectral signal-to-noise ratio (SNR) suitable for resolving subtle differences in the EELS fine structure. 

Particularly for the O-K edge, acquisitions up to 2000 sec for a single $\sim$10-20 nm$^2$ region were needed in order to record spectra with sufficiently high SNR. To ensure low spatial drift over such long measurements, EELS experiments of the O-K edge were performed on a Nion UltraSTEM equipped with a high stability stage, an Enfinium ER spectrometer, and Quefina 2 camera, operated at 100 kV with 12 pA beam current. The effective energy resolution measured by the full width at half-maximum (FWHM) of the zero loss peak (ZLP) was $\sim$0.39 eV. In each specimen, atomic-resolution STEM imaging was used to locate defect-free regions and eliminate contributions from SrTiO$_3$ substrate and capping layers. 

Most spectra presented here are the sum of multiple acquisitions. All individual spectra were aligned in energy based on cross-correlation of the large spectral features. In some cases (Ni-L$_{2,3}$ and Nd-M$_{4,5}$ edges from the NdNiO$_3$ film, main text Fig. 2), a weighted moving average over a small window (3-5 channels) has been applied for clarity of comparison between spectra. The energy dispersion of 0.025 eV/ch is significantly smaller than the instrumental energy resolution, so smoothing to this degree will not affect the overall interpretation of our data. Absolute energy alignment of the O-K edges of the Nd$_{1-x}$Sr$_x$NiO$_y$ thin films are based on simultaneously recorded O-K spectra from the SrTiO$_3$ substrates, which serves as a consistent and well-documented reference. In each figure, spectra are normalized and shifted vertically for clarity and ease of comparison. 

To obtain sufficient SNR of the high-energy edges (Ni-L$_{2,3}$, Nd-M$_{4,5}$), we took advantage of the low background and readout noise in the Gatan K2 Summit direct electron detector installed on the FEI Titan Themis system  \cite{hart_direct_2017, goodge_direct_2018}. The inherent energy spread of the X-FEG electron source on this system, however, required monochromation to obtain sufficient energy resolution. For these experiments we used a monochromator excitation of 1.0 at an accelerating voltage of 120 kV to achieve measured energy resolution of $\sim$0.21 eV with a spectrometer dispersion of 0.025 eV/ch. Because of the point spread function (PSF) of the K2 Summit detector \cite{hart_direct_2017, goodge_direct_2018}, the 0.1 eV/ch dispersion used for the experiments preserved an effective energy resolution of $\sim$0.29 eV over a total energy range of 370 eV, enabling simultaneous acquisition of both the Ni-L$_{2,3}$ and the Nd-M$_{4,5}$ edge used for absolute energy alignment (Fig. S4). 
Ni and Nd spectra were acquired by the same process described for the O-K edge. The O-K signature was checked for each region before and after additional data collection to ensure proper phase identification (i.e., perovskite vs. infinite-layer) and to confirm that the results were not affected by radiation damage during the experiment (Fig. S7). All spectra were aligned in energy based on a simultaneously recorded nearby standard edge: e.g., the Ni-L$_{2,3}$ spectra were aligned by the Nd-M$_{4,5}$, the Nd-N$_{2,3}$ by the C-K edge.

\subsection*{Acknowledgements}
B.H.G. and L.F.K. acknowledge support by the Department of Defense Air Force Office of Scientific Research (No. FA 9550-16-1-0305). This work made use of the Cornell Center for Materials Research (CCMR) Shared Facilities, which are supported through the NSF MRSEC Program (No. DMR-1719875). The FEI Titan Themis 300 was acquired through No. NSF-MRI-1429155, with additional support from Cornell University, the Weill Institute, and the Kavli Institute at Cornell. The Thermo Fisher Helios G4 X FIB was acquired with support from the National Science Foundation Platform for Accelerated Realization, Analysis, and Discovery of Interface Materials (PARADIM) under Cooperative Agreement No. DMR-1539918. The work at SLAC/Stanford is supported by the US Department of Energy, Office of Basic Energy Sciences, Division of Materials Sciences and Engineering, under contract number DE-AC02-76SF00515; and the Gordon and Betty Moore Foundation’s Emergent Phenomena in Quantum Systems Initiative through grant number GBMF4415 (synthesis equipment). M.O. acknowledges partial financial support from the Takenaka Scholarship Foundation. G.A.S. acknowledges support by the Natural Sciences and Engineering Research Council (NSERC) and the Stewart Blusson Quantum Matter Institute of UBC.


\end{document}